\documentclass[sn-mathphys,Numbered,iicol]{sn-jnl}

\usepackage{graphicx}%
\usepackage{multirow}%
\usepackage{amsmath,amssymb,amsfonts}%
\usepackage{amsthm}%
\usepackage{mathrsfs}%
\usepackage[title]{appendix}%
\usepackage{xcolor}%
\usepackage{textcomp}%
\usepackage{manyfoot}%
\usepackage{booktabs}%
\usepackage{algorithm}%
\usepackage{algorithmicx}%
\usepackage{algpseudocode}%
\usepackage{listings}%

\theoremstyle{thmstyleone}%
\theoremstyle{thmstyletwo}%

\theoremstyle{thmstylethree}%

\begin{document}

\title[]{Experimental and theoretical total cross sections for single and double ionization of the open-$4d$-shell ions Xe$^{12+}$, Xe$^{13+}$, and Xe$^{14+}$ by electron impact\\ {\small \today}}


\author*[1,2]{\fnm{Fengtao} \sur{Jin}}\email{ftjin@nudt.edu.cn}

\author[1]{\fnm{Alexander} \sur{Borovik, Jr.}}

\author[1,3]{\fnm{B. Michel} \sur{D{\"o}hring}}

\author[1]{\fnm{Benjamin} \sur{Ebinger}}

\author[1]{\fnm{Alfred} \sur{M{\"u}ller}}

\author*[1,3]{\fnm{Stefan} \sur{Schippers}}\email{schippers@jlug.de}

\affil[1]{\orgdiv{I. Physikalisches Institut}, \orgname{Justus-Liebig Universt{\"a}t Gie{\ss}en}, \orgaddress{\postcode{35392} \city{Giessen}, \country{Germany}}}

\affil[2]{\orgdiv{Department of Physics}, \orgname{National University of Defense Technology}, \orgaddress{\postcode{410073} \city{Changsha}, \country{China}}}

\affil[3]{\orgdiv{Helmholtz Forschungsakademie Hessen f\"ur FAIR (HFHF), GSI Gesellschaft f\"ur Schwerionenforschung, Campus Gie{\ss}en}, \orgaddress{\postcode{35392} \city{Giessen}, \country{Germany}}}




\abstract{We present new experimental and theoretical cross sections for electron-impact single ionization of Xe$^{12+}$ and Xe$^{13+}$ ions, and double ionization of Xe$^{12+}$, Xe$^{13+}$ and Xe$^{14+}$ ions for collision energies from the respective ionization thresholds up to 3500 eV. The calculations use the fully relativistic subconfiguration-averaged distorted-wave (SCADW) approach and, partly, the more detailed level-to-level distorted wave (LLDW) method. We find that, unlike in previous work, our theoretical cross sections agree with our experimental ones within the experimental uncertainties, except for the near-threshold double-ionization cross sections. We attribute this remaining discrepancy to the neglect of direct-double ionization in the present theoretical treatment.
}

\maketitle
\section{Introduction}\label{intro}

Electron-impact ionization (EII) is one of the fundamental atomic collision processes that governs the charge balance in hot plasmas \cite{Mueller2008a}. Prominent examples of research fields that require large amounts of atomic collision cross sections are astrophysics \cite{Kallman2007a,Savin2012} and magnetically confined nuclear fusion \cite{Mueller2015b,Murakami2015,Puetterich2019}. In these applications cross sections not only for single ionization but also for multiple ionization are of much interest \cite{Hahn2017a}.

Here, we provide new experimental and theoretical data for electron-impact single ionization (EISI) and electron-impact double ionization (EIDI) of multiply charged xenon ions, i.e., Xe$^{12+}$, Xe$^{13+}$, and Xe$^{14+}$, which all have an open $4d$ subshell. Such many-electron open-shell systems pose a formidable challenge for atomic theory. Another challenging aspect of particularly electron-impact ionization arises from the fact that the ionization event results in at least two electrons in the continuum which can mutually interact with each other in the field of the (generally polarizable) product ion.  

In recent years, several studies on electron-impact ionization of open-$4d$-subshell xenon ions (Xe$^{q+}$ with $9\leq q \leq 17$) have been conducted. \citet{Borovik2011} presented experimental and theoretical results for EISI of Xe$^{10+}$ ions in the experimental energy range from below the ground-level single ionization threshold at 229~eV up to 5000 eV with the experimental data beyond 1000~eV having been taken from earlier work by \citet{Hofmann1993}, who also measured EISI cross sections for Xe$^{9+}$. The theoretical results of \citet{Borovik2011} for EISI of Xe$^{10+}$ were obtained with the nonrelativistic configuration-average distorted-wave (CADW) method and agreed within the experimental uncertainty with the measured EISI cross section over the entire experimental energy range.

\citet{Pindzola2013} presented a joint experimental and theoretical study on EISI of  moderately charged xenon ions ($q=10-17$). The experimental energy range of 200-1000~eV comprised all single-ionization thresholds of these ions. The calculations with the CADW method considered direct ionization and indirect excitation-autoionization processes for ground-configuration ions and metastable excited-configuration ions.  The resulting theoretical cross sections deviated significantly from the measured ones, particularly for ion charge states $q\geq 11$. Later, more refined CADW calculations by \citet{Borovik2015} for $q=8-17$ improved the agreement of theory and experiment but did not significantly change the overall picture. Nevertheless, the computational results were very useful for providing experimentally-derived EISI plasma rate coefficients. 

The present investigation extends the above mentioned previous experimental work \cite{Pindzola2013,Borovik2015} to collision energies of up to 3500~eV and also addresses EIDI in addition to EISI. Our experimental method is described in Sec.~\ref{sec:exp}. Also the present theoretical approach is different from the previous nonrelativistic CADW calculations. We used two different variants of distorted-wave calculations, i.e., the relativistic subconfiguration-averaged distorted-wave (SCADW) method \cite{Pindzola1988} and the even more elaborate level-to-level distorted-wave (LLDW) method \cite{Griffin1982}. Our theoretical approaches are detailed in Sec.~\ref{sec:theo}. Our experimental and theoretical results are jointly presented and discussed in Sec.~\ref{result}. The concluding Sec.~\ref{sec:conc} provides a short summary of the present key findings as well as a brief outlook on possible future research directions.

\section{Experimental approach}\label{sec:exp}

The experimental data were measured at the Giessen electron-ion crossed-beams setup \cite{Jacobi2004a,Rausch2011}. In this setup, multiply charged xenon ions were produced in an electron cyclotron resonance (ECR) ion source \cite{Broetz2001} where oxygen was added to the xenon plasma in order to maximize the production of Xe$^{12+}$, Xe$^{13+}$, and Xe$^{14+}$ ions as described already earlier \cite{Pindzola2013}. The ions were accelerated by putting the ion source on a high voltage  of 12~kV, and the desired ion species was selected according to its mass-to-charge ratio by passing the ion beam through a combination of narrow slits and a double-focussing dipole magnet. The selected ions were then collimated to a size of about 1~mm~$\times$~1~mm by two sets of four-jaw slits and subsequently crossed with an electron beam under an angle of 90$^\circ$. Product ions resulting from electron-impact ionization (and from collisions with residual-gas particles) were separated from the primary ion beam by a second magnet. Ions in the chosen product charge state were counted with nearly 100\% efficiency on a channeltron-based single particle detector \cite{Rinn1982}. The primary ions were collected in a Faraday cup and the ion current was recorded with a sensitive electrometer. For the ions under study, the electrical ion currents in the electron-ion interaction region were approx.\ 8.5~nA for $^{129}$Xe$^{12+}$, 3.6~nA for $^{129}$Xe$^{13+}$, and 3.5~nA for $^{129}$Xe$^{14+}$.

The electron beam was provided by an electron gun which is different from the gun used in our earlier work \cite{Pindzola2013,Borovik2015}. In particular, the new gun extends the available electron-energy range from previously maximally 1000~eV to now 3500~eV \cite{Shi2003a,Ebinger2017,Doehring2020}. The electron gun produces a ribbon-shaped electron beam with currents of up to 900 mA at the maximum energy. For a specific measurement, however, the electron current depends on the electron energy and the selected operation mode of the electron gun \cite{Ebinger2017}. 

For the determination of absolute EII cross sections the mutual geometrical overlap of both particle beams was measured by moving the electron beam mechanically through the ion beam and by simultaneously recording the ionization signal as a function of gun position. This animated crossed-beams technique \cite{Defrance1981,Mueller1985,Mueller1985a} also quantifies the background resulting from ionization in collisions of the primary ions with residual-gas particles such that it can be readily subtracted. Another source of background were ions which were trapped in the space charge potential of the electron beam. This background was eliminated by a space-charge compensation technique \cite{Mueller1987a}, which involved the leaking of xenon or nitrogen gas into the interaction chamber. Thereby, the vacuum pressure rose from a background pressure of some 10$^{-10}$~mbar to about 10$^{-7}$~mbar.

The systematic uncertainties of the absolute cross sections were estimated as the quadrature sum of the uncertainties of all parameters used for calculating the cross section. This approach is very similar to the one used at this setup with the former electron gun still in operation \cite{Rausch2011} and only the estimation of the electron current's uncertainty was adjusted. It results in systematic uncertainties of the measured EII cross sections between 6.5\% and 7\% depending on the particular electron beam's characteristics for every single measurement. For energies well above the particular ionization threshold, statistical uncertainties are typically less than 1\% at 95\% confidence level resulting in a dominance of the systematic uncertainties when evaluating the total uncertainties. Only close to the ionization threshold the statistical uncertainties and thus also the total uncertainties are higher due to a low ionization signal. Also for double ionization of Xe$^{14+}$, the total uncertainties were larger for all measurements above 2500 eV because of an accidentally incomplete compensation of the electron beam's space-charge for these measurements. All error bars in the figures below correspond to the quadrature sum of the statistical and systematic uncertainties.

Although the contribution of the ion energy to the electron-ion collision energy is almost negligible, it was properly accounted for in the nonrelativistic calculation of the latter. The uncertainty of the electron-ion collision energy scale was about 10~eV.

\section{Theoretical method}\label{sec:theo}


\begin{table}[]
	\caption{Single ionization potentials (SIP) and double-ionization potentials (DIP) of ground-level Xe$^{12+}$, Xe$^{13+}$, and Xe$^{14+}$ ions in eV: Present  FAC results and values from the NIST Atomic Spectra database \cite{Kramida2022a}. Both sets of data agree within the quoted uncertainties of the NIST values, keeping in mind that the FAC values also have an (unspecified) uncertainty.}
	\label{tab:thres}
	\begin{tabular}{@{}lllllll@{}}
		\toprule
		& \multicolumn{2}{c}{Xe$^{12+}$}                          & \multicolumn{2}{c}{Xe$^{13+}$}                          & \multicolumn{2}{c}{Xe$^{14+}$}                          \\ 
		& \multicolumn{1}{c}{SIP} & \multicolumn{1}{c}{DIP} & \multicolumn{1}{c}{SIP} & \multicolumn{1}{c}{DIP} & \multicolumn{1}{c}{SIP} & \multicolumn{1}{c}{DIP} \\ \midrule
		FAC  & 282.38                  & 592.79                  & 310.41                  & 653.47                  & 343.06                  & 715.66                  \\
		NIST & 281(3)                  & 595(4)                  & 314(3)                  & 657(4)                  & 343(3)                  & 717(4)                  \\ \bottomrule
	\end{tabular}
\end{table}

We used quantum mechanical perturbation theory for the computation of cross sections for electron-impact single and double ionization of Xe$^{12+}$, Xe$^{13+}$, and Xe$^{14+}$ ions at electron energies ranging from the individual ionization thresholds (Tab.~\ref{tab:thres}) up to 4000~eV. All required atomic quantities were obtained from the Flexible Atomic Code (FAC) \cite{Gu2008},  which is an implementation of the fully relativistic Dirac-Fock-Slater method for the calculation of atomic energy levels and wave functions (see also \cite{Sampson2009}). These wave functions were used for computing radiative and autoionizing atomic transition rates involving both bound and continuum levels. Within the current approach, the continuum wave functions are referred to as distorted waves (DW). 

The computational effort depends on the level of detail requested. Depending on the individual configurations involved, we carried out two different variants of DW calculations, i.e., the subconfiguration-averaged distorted-wave (SCADW) method \cite{Pindzola1988} and the level-to-level distorted-wave (LLDW) method \cite{Griffin1982}. We have applied this hybrid approach already earlier to EII of W$^{14+}$, W$^{15+}$, and W$^{16+}$ ions \cite{Jin2020,Jin2020a}, where we have shown that the much more costly LLDW method becomes only necessary when the less demanding SCADW method averages autoionizing and non-autoionizing levels to joint subconfigurations that are either autoionizing or non-autoionizing as a whole and, thus, cannot represent the true atomic structure correctly. As will be detailed below, this is also the case for some of the configurations included in the present calculations.

\subsection{Electron-impact single ionization}

There are several processes that result in electron-impact single-ionization (EISI) of ions. The simplest process is direct ionization (DI) where an initially bound electron is directly removed from an ion through a collision with an incoming electron. The DI process can be formally described by
\begin{eqnarray}
	e^{-}+A^{q+}\rightarrow 2e^{-}+A^{(q+1)+}.
\end{eqnarray}

By electron-impact, the ion can also be excited to an autoionizing level and then autoionize to the next higher charge state via an Auger process. This so-called excitation-autoionization (EA) process 
can be viewed as a sequence of two steps, i.e.
\begin{eqnarray}
	e^{-}+A^{q+}\rightarrow e^{-}+[A^{q+}]^{*}\rightarrow 2e^{-}+A^{(q+1)+},
\end{eqnarray}
where $[A^{q+}]^{*}$ denotes the intermediately excited ion. The EA process contributes significantly to the total EISI cross section, especially near the ionization threshold and for moderately and highly charged ions. 

In some situations, also the resonant-excitation double-autoionization (REDA) and the resonant-excitation auto-double ionization (READI) processes can be important \cite{Mueller2008a}. In these two processes, the $A^{q+}$ ion first captures an incident electron by dielectronic capture (DC) such that an excited level in the lower charged ion $A^{(q-1)+}$ is generated. If this level is above the double-ionization potential of this ion it autoionizes by emitting with a certain probability \emph{two} electrons either sequentially (REDA) or simultaneously (READI) such that a net ionization has occurred. For REDA, this is formally described as
\begin{eqnarray}
	e^{-}+A^{q+}\rightarrow [A^{(q-1)+}]^{**} \rightarrow e^{-}+[A^{q+}]^{*} \nonumber \\
	\rightarrow 2e^{-}+A^{(q+1)+}.\label{eq:REDA}
\end{eqnarray}
REDA has been found to make significant contributions to EISI of Xe$^{24+}$ ions \cite{Liu2015}. In contrast to REDA, which is governed by two-electron interactions, the READI process involves a much less probable three-electron interaction. Accordingly, its cross section is usually smaller than that for REDA. Therefore, READI and more higher-order processes are disregarded in the present study.

The total EISI cross section from level $i$ of an ion A$^{q+}$ to level $j$ of the ionized ion A$^{(q+1)+}$ can be expressed as
\begin{eqnarray}
	&\sigma_{ij}(\varepsilon)&=\sigma_{ij}^{DI}(\varepsilon)+\sum_k{\sigma_{ik}^{CE}(\varepsilon)B_{kj}} \nonumber \\
	& &+\sum_{lm}{\sigma_{il}^{DC}(\varepsilon)B_{lm}B_{mj}}.
	\label{eq:EISI}
\end{eqnarray}
Here, $\sigma_{ij}^{DI}$ is the DI cross section at the incident electron energy $\varepsilon$, and $\sigma_{ik}^{CE}$ is the electron-impact collisional-excitation cross section from level $i$ to an autoionizing level $k$ of the ion. $\sigma_{il}^{DC}$ is the DC cross section from the level $i$ of A$^{q+}$ to the level $l$ of the lower charge state ion A$^{(q-1)+}$ with the requirement that level $l$ needs to be capable of double autoionization. $B_{kj}$, $B_{lm}$, and $B_{mj}$ represent the branching ratios (BR) for autoionization. $B_{kj}$ is the BR of the EA process, which is determined by the expression
\begin{eqnarray}
	B_{kj}=\frac{A_{kj}^{a}+\sum_n{A^r_{kn}B_{nj}}}{\sum_m{A_{km}^{a}+\sum_n{A_{kn}^{r}}}},
	\label{eq:BR}
\end{eqnarray}
where $A_{kj}^a$ is the Auger rate from level $k$ of the ion $A^{q+}$ to level $j$ of the ion $A^{(q+1)+}$, and $A^r$ is the radiative transition probability. The second term in the numerator accounts for radiative transitions into energetically lower autoionizing levels $n$ of the ion $A^{q+}$. In  Eq.~\ref{eq:EISI}, $B_{lm}$ and $B_{mj}$ are defined in the analogous way for the two autoionization steps required for a REDA process from level $i$ of ion $A^{q+}$ to level $j$ of the ionized ion $A^{(q+1)+}$ via the intermediate levels $l$ and $m$. 

In the present EISI calculations, contributions from  DI of $K$-, $L$-, and $M$-shell electrons are not considered because all the resulting hole states can --- and in the vast majority of cases will --- result in at least another Auger process and effectively lead to net multiple ionization of the original ion. Therefore, only DI of the 4$s$, $4p$ and $4d$ subshells has been taken into account in the present calculations.  Concretely, the considered DI channels are
\begin{eqnarray}
	e+\text{[Ar]}3d^{10}4s^{2}4p^{6}4d^{6-x}  
	\rightarrow 2e+\left\lbrace 
	\begin{array}{l}
		4s^{2}4p^{6}4d^{5-x}\\
		4s^{2}4p^{5}4d^{6-x}\\
		4s^{1}4p^{6}4d^{6-x},
	\end{array}
	\right.
	\label{eq:DI}
\end{eqnarray}
where $x=q-12=0,1,2$ for the ion charge states $q=12,13,14$, respectively.

In the calculation of the EA cross section all excitations that result in states with the potential to autoionize must be included. This applies in particular to the excitation of $M$-shell electrons. In analogy to the DI, the excitations of electrons from the $K$ and $L$ shells are not considered as these will result in effective multiple ionization. The EA channels that have been taken into account in our calculations are
\begin{eqnarray}
	e&+&\text{[Ne]}3s^{2}3p^{6}3d^{10}4s^{2}4p^{6}4d^{6-x} \nonumber \\
	& &\rightarrow e+\left\{ 
	\begin{array}{l}
		3s^{2}3p^{6}3d^{10}4s^{2}4p^{5}4d^{6-x}nl\\
		3s^{2}3p^{6}3d^{10}4s^{1}4p^{6}4d^{6-x}nl \\
		3s^{2}3p^{6}3d^{9\phantom{1}}4s^{2}4p^{6}4d^{6-x}nl \\
		3s^{2}3p^{5}3d^{10}4s^{2}4p^{6}4d^{6-x}nl \\
		3s^{1}3p^{6}3d^{10}4s^{2}4p^{6}4d^{6-x}nl,
	\end{array}
	\right.
	\label{eq:EA}
\end{eqnarray}
where all $n\leq25$ and $l\leq8$ had to be considered for reaching convergence of the total EA cross sections. It should be noted that $nl$ also includes the $4d$ subshell.

For the calculation of the REDA cross section -- analogous to the EA -- all DC channels that result in states with the potential for double-ionization but not triple-ionization need to be included, i.e. 
\begin{eqnarray}
	e&+&\text{[Ne]}3s^{2}3p^{6}3d^{10}4s^{2}4p^{6}4d^{6-x} \nonumber \\
	& &\rightarrow \left\{ 
	\begin{array}{l}
		3s^{2}3p^{6}3d^{10}4s^{2}4p^{5}4d^{6-x}n'l'nl\\
		3s^{2}3p^{6}3d^{10}4s^{1}4p^{6}4d^{6-x}n'l'nl\\
		3s^{2}3p^{6}3d^{9\phantom{1}}4s^{2}4p^{6}4d^{6-x}n'l'nl \\
		3s^{2}3p^{5}3d^{10}4s^{2}4p^{6}4d^{6-x}n'l'nl \\
		3s^{1}3p^{6}3d^{10}4s^{2}4p^{6}4d^{6-x}n'l'nl,
	\end{array}
	\label{eq:DC}
	\right.
\end{eqnarray}
where $n'\leq 9$, $l'\leq 5$ ($n'l'$ also refers to $4d$), and $n\leq 25$, $l\leq 5$. Analogous to EA, these ranges of quantum numbers of the excited electrons proved to be sufficiently high for the total REDA cross sections to converge. For the decay of the thus formed $[A^{(q-1)+}]^{**}$ ions (cf.~Eq.~\ref{eq:REDA}), all possible Auger and radiative decay channels were included to obtain the autoionization branching ratios (Eq.~\ref{eq:BR}) that are needed for the computation of the various partial cross sections contributing to Eq.~\ref{eq:EISI}. 

\subsection{Electron-impact double ionization}

The conceptually simplest contribution to EIDI is direct-double ionization (DDI)
\begin{equation}\label{eq:DDI}
	e^- + A^{q+} \to 3e^- + A^{(q+2)+},
\end{equation}
where two initially bound electrons are simultaneously released. This can happen by two collisions of the incident electron with two bound electrons, or by one collision of the projectile with one bound electron which then knocks off another bound electron on its way out. A third possibility is that the projectile electron knocks out a bound electron and by the sudden change of screening a second bound electron is shaken off.

A theoretical description of this double-ionization channel is notoriously difficult because it requires a thorough treatment of the correlations between the three outgoing continuum electrons \cite{Berakdar2003}. To date, a general computational method for the calculation of DDI cross sections is not available, and the development of such a method is beyond the scope of the present work. Therefore, DDI is neglected in our present EIDI calculations. 

Similarly to what has been done in recent work by others \cite{Konceviciute2021,Ma2023} we take only indirect ionization processes into account. In particular, we consider ionization-autoionization (IA) and excitation double-autoionization (EDA), which are the dominant contributions to EIDI of heavy many-electron ions. In IA an inner-shell electron of the $A^{q+}$ ion is initially removed by direct ionization such that a multiply excited state of the ionized $A^{(q+1)+}$ ion is formed, which subsequently decays via autoionization resulting in the production of a net doubly ionized $A^{(q+2)+}$ ion. In EDA the $A^{q+}$ ion is excited to a level in the double-ionization continuum, which autoionizes in two consecutive steps to form an $A^{(q+2)+}$ ion. In principle the autoionization can also occur in one step, where two electrons are emitted simultaneously and which is called excitation auto-double ionization (EADI). As already discussed for REDA and READI, such higher-order processes typically are of minor importance and were neglected as well to keep the calculations tractable.

The total double-ionization cross section resulting from IA and EDA  can be written as
\begin{equation}\label{eq:EIDI}
	\sigma_{ij}(\varepsilon)=\sum_k{\sigma_{ik}^{DI}(\varepsilon)B_{kj}}+\sum_{lm}{\sigma_{il}^{CE}(\varepsilon)B_{lm}B_{mj}},
\end{equation}
where the subscripts $i$ and $j$ refer to an initial level in the primary $A^{q+}$ ion and a final level in the product ion $A^{(q+2)+}$, respectively. The quantity $\sigma_{ik}^{DI}$ is the direct ionization cross section from level $i$ of $A^{q+}$ to the autoionizing level $k$ of $A^{(q+1)+}$, and $B_{kj}$ is the autoionization branching ratio from level $k$ to level $j$ of $A^{(q+2)+}$, and $\sigma_{il}^{CE}$ is the electron-impact excitation cross section from level $i$ to level $l$ of $A^{q+}$, and $B_{lm}$ and $B_{mj}$ are the autoionization branching ratios from level $l$ to the autoionizing level $m$ of $A^{(q+1)+}$ and from level $m$ to the non-autoionizing level $j$ of $A^{(q+2)+}$, respectively.

In the present calculations, we considered the IA channels 
\begin{eqnarray}
	e&+&\text{[Ne]}3s^{2}3p^{6}3d^{10}4s^{2}4p^{6}4d^{6-x}  \nonumber \\
	& &\rightarrow \left\{
	\begin{array}{l}
		3s^{2}3p^{6}3d^{9}4s^{2}4p^{6}4d^{6-x} \\
		3s^{2}3p^{5}3d^{10}4s^{2}4p^{6}4d^{6-x} \\
		3s^{1}3p^{6}3d^{10}4s^{2}4p^{6}4d^{6-x}, \\
	\end{array}
	\label{eq:IA}
	\right.
\end{eqnarray}
and the EDA channels
\begin{eqnarray}
	e&+&\text{[Ne]}3s^{2}3p^{6}3d^{10}4s^{2}4p^{6}4d^{6-x} \nonumber \\
	& &\rightarrow e+\left\{ 
	\begin{array}{l}
		3s^{2}3p^{6}3d^{9}4s^{2}4p^{6}4d^{6-x}nl \\
		3s^{2}3p^{5}3d^{10}4s^{2}4p^{6}4d^{6-x}nl \\
		3s^{1}3p^{6}3d^{10}4s^{2}4p^{6}4d^{6-x}nl.
	\end{array}
	\right.
	\label{ew:EAA}
\end{eqnarray}
For both, IA and EDA, calculations were performed for $4 \leq n \leq 25$ and $l\leq6$ as required for the convergence of the total cross section.

\subsection{Computations for metastable primary ions}

Up to now, we have implicitly assumed that all ions were initially in their respective ground configurations. However, the ECR ion source also produces excited ions. If the excited levels are sufficiently long-lived, they will survive their transport from the ion source to the  electron-ion collision region. In the present experiments, the corresponding flight times were about 9~$\mu$s at an acceleration voltage of 12 kV \cite{Borovik2015}. 

In the measured cross sections, metastable ions reveal their presence by producing an ionization signal at energies below the threshold for ionization of the ground level. In the EISI cross sections of all ions under consideration, there are weak contributions of that kind, which suggests that minor fractions of the primary ions were initially in excited metastable levels which, according to comprehensive lifetime calculations, belong to excited levels of the [Kr]$4d^{6-x-1}4f$ excited configuration \cite{Borovik2015}. Therefore, we performed cross section calculations also for EISI and EIDI of ions in initially [Kr]$4d^{6-x-1}4f$ configurations, which were carried out analogous to the calculations for ground state ions. As discussed in detail below, the (small) metastable ion fractions were finally inferred from the comparison of the calculated total EISI cross sections with the measured ones.

\section{\label{result}Results and discussions}

\subsection{Electron impact single ionization}

\begin{figure}
	\includegraphics[width=\columnwidth]{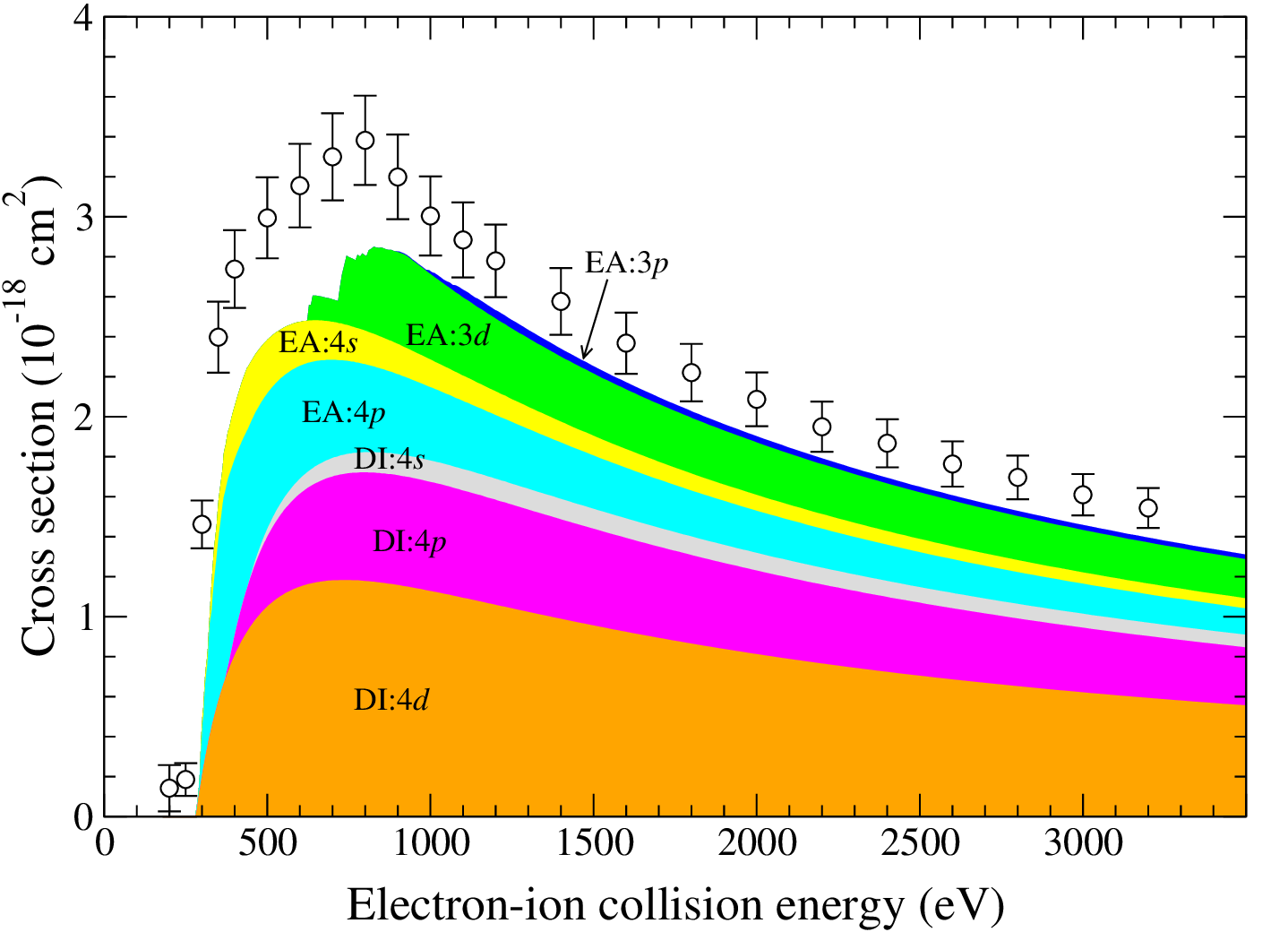}
	\caption{\label{xe12eisi-scadw}EISI cross section of Xe$^{12+}$: Present experimental results (open circles with error bars) and purely SCADW results for Xe$^{12+}$ in its ground-subconfiguration with the various shaded curves representing the indicated contributions to the total cross section.}
\end{figure}

Figure \ref{xe12eisi-scadw} shows our experimental and theoretical SCADW results for the EISI cross section of Xe$^{12+}$. The experimental results agree with our earlier ones \cite{Pindzola2013,Borovik2015} (not shown), which were limited to a maximum collision energy of 1000~eV. This is also the case for the other two ions presently under study. The SCADW calculations comprised DI and EA and only the ground configuration [Ar]$3d^{10}4s^{2}4p^{6}4d^{6}$ was considered. According to our calculations and in agreement with the results from our earlier CADW calculations \cite{Borovik2015}, the largest contributions to the EISI cross section in the experimental energy range are by DI of the $4d$ and $4p$ subshells and by EA involving the excitation of a $4p$ or a $3d$ electron. The agreement between experiment and theory is not totally satisfying. The theoretical cross section is significantly smaller than the experimental one, particularly, below the cross-section maximum at about 800~eV. This discrepancy is due to  oversimplifications in the model used for calculating the cross section in Fig.~\ref{xe12eisi-scadw}. In the following we refine our theoretical model by partly applying the fine-structure resolving LLDW approach and by accounting for REDA processes as well for the presence of metastable primary ions.

\begin{figure}
	\includegraphics[width=\columnwidth]{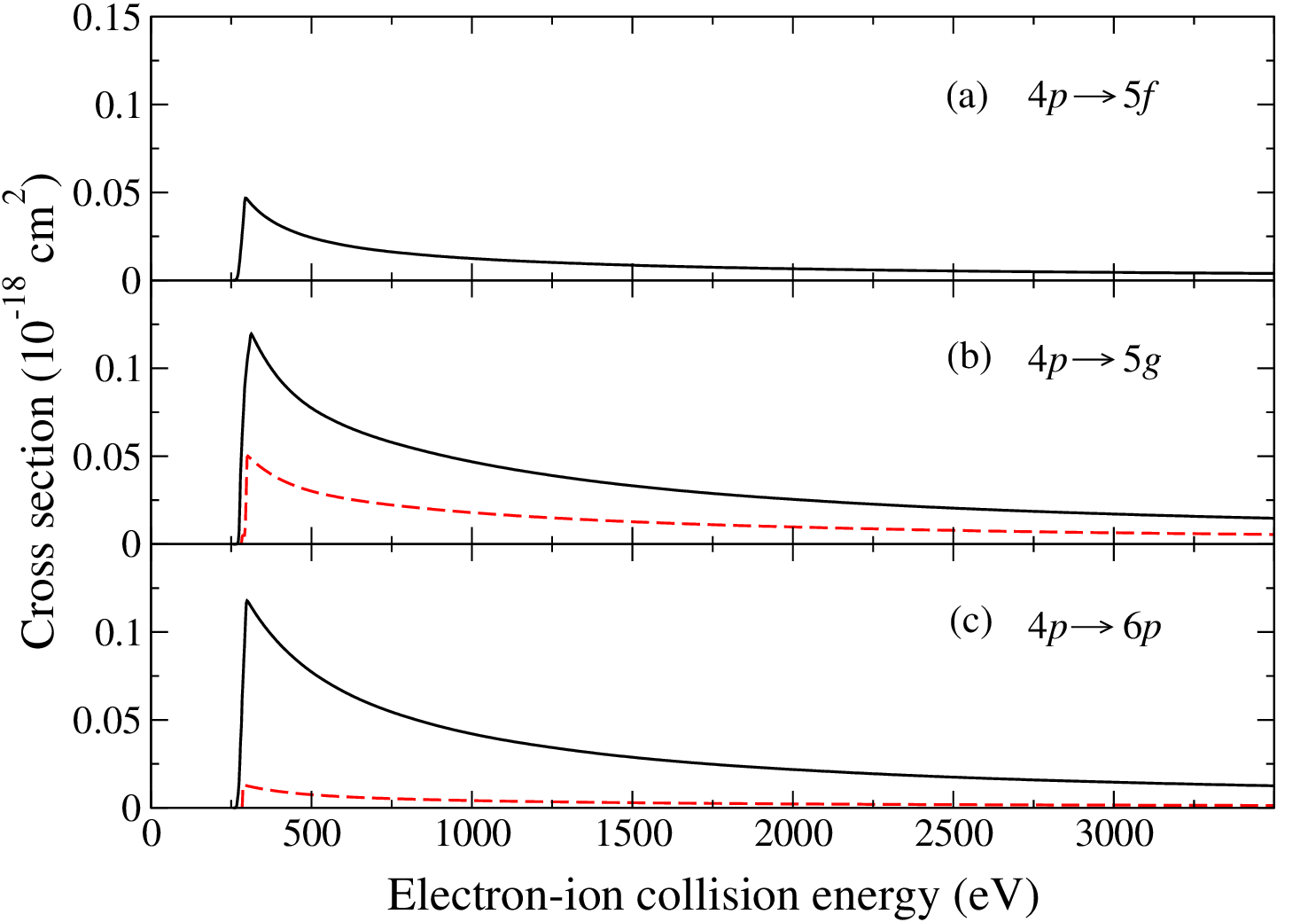}
	\caption{\label{ea4p-patch} Comparisons between LLDW (solid black lines) and SCADW (dashed red lines) EA cross sections for the $4p\to 5f$ (panel~a), $4p\to 5g$ (panel~b), and $4p\to 6p$ (panel~c) excitation channels.}
\end{figure}

As discussed already in section \ref{sec:theo}, the EISI cross section cannot be calculated correctly by the SCDAW  model whenever it combines  autoionizing and non-autoionizing levels to joint subconfigurations that are either autoionizing or non-autoionizing as a whole. This affects predominantly the EA cross sections with the excitation energies being in the vicinity of the ionization threshold. Examples are shown in Fig.~\ref{ea4p-patch} which compares LLDW and SCADW EA cross sections for some $4p$ excitation channels. The results of both methods are strikingly different from each other.

In the SCADW calculation, the $4p\rightarrow 5f$ excitation to the $4s^{2}4p^{5}4d^{6}5f$ configuration splits into 20 relativistic subconfigurations spreading out over excitation energies ranging from  260 to 281~eV, which are all below the ionization threshold. Therefore, this channel does contribute to EA in the SCADW model. In the LLDW calculation the situation is much different. There the $4s^{2}4p^{5}4d^{6}5f$ configuration splits into 2022 levels which are distributed over the energy range $251 - 306$~eV. Among these levels are 688 above the ionization threshold which thus contribute to EA. A similar reasoning also applies to the $4p\to5g$ and $4p\to6p$ excitation channels where LLDW calculations predict a larger number of above-threshold contributions than the SCADW method. In all cases this leads to significantly larger LLDW EA cross sections as compared to the corresponding SCADW cross sections, which emphasizes the need for a detailed level-to-level treatment of the excitation channels that straddle the threshold. 

\begin{figure}
	\includegraphics[width=\columnwidth]{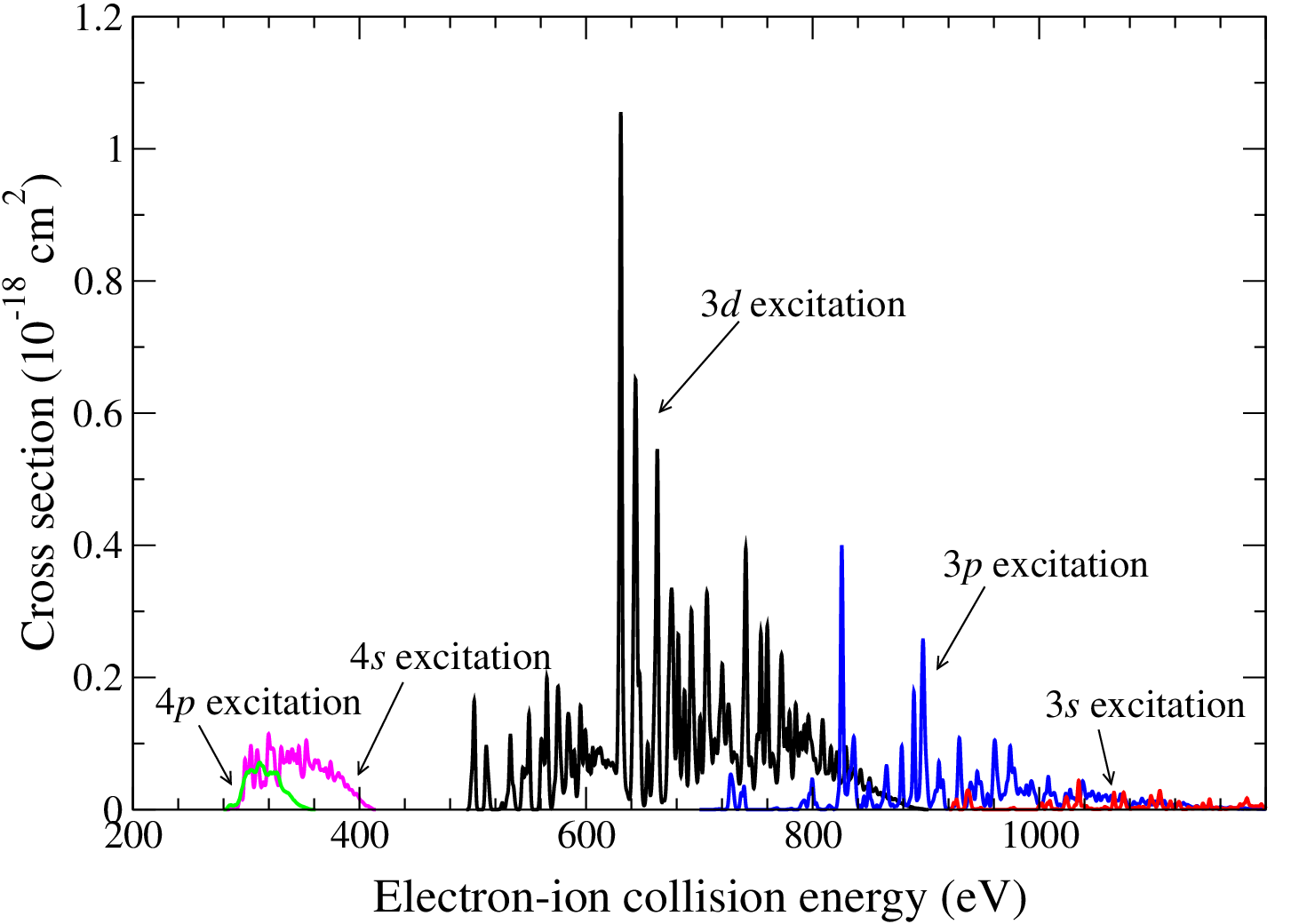}
	\caption{\label{xe12-reda} REDA of Xe$^{12+}$. The black line refers to the REDA in which in the dielectronic capture process a $3d$ electron is excited. The blue, red, magenta and green lines refer to the $3p$, $3s$, $4s$ and $4p$ excitations, respectively.}
\end{figure}

Another improvement of our calculations concerns the inclusion of REDA (Eq.~\ref{eq:REDA}) in the theoretical EISI cross sections. REDA was not treated in our earlier work \cite{Borovik2015}. Figure~\ref{xe12-reda} shows the computed REDA cross sections for Xe$^{12+}$. Their contribution to the total EISI cross section (Fig.~\ref{xe12eisi-scadw}) is significant. The strongest REDA channels involve the excitation of a $3d$ electron. The open $3d$ subshell leads to a large number of doubly excited levels which span the energy range $500-900$~eV, i.e., to beyond the maximum of the total EISI cross section where also resonances from REDA processes involving the excitation of a $3p$ electron occur.

\begin{figure}
	\includegraphics[width=\columnwidth]{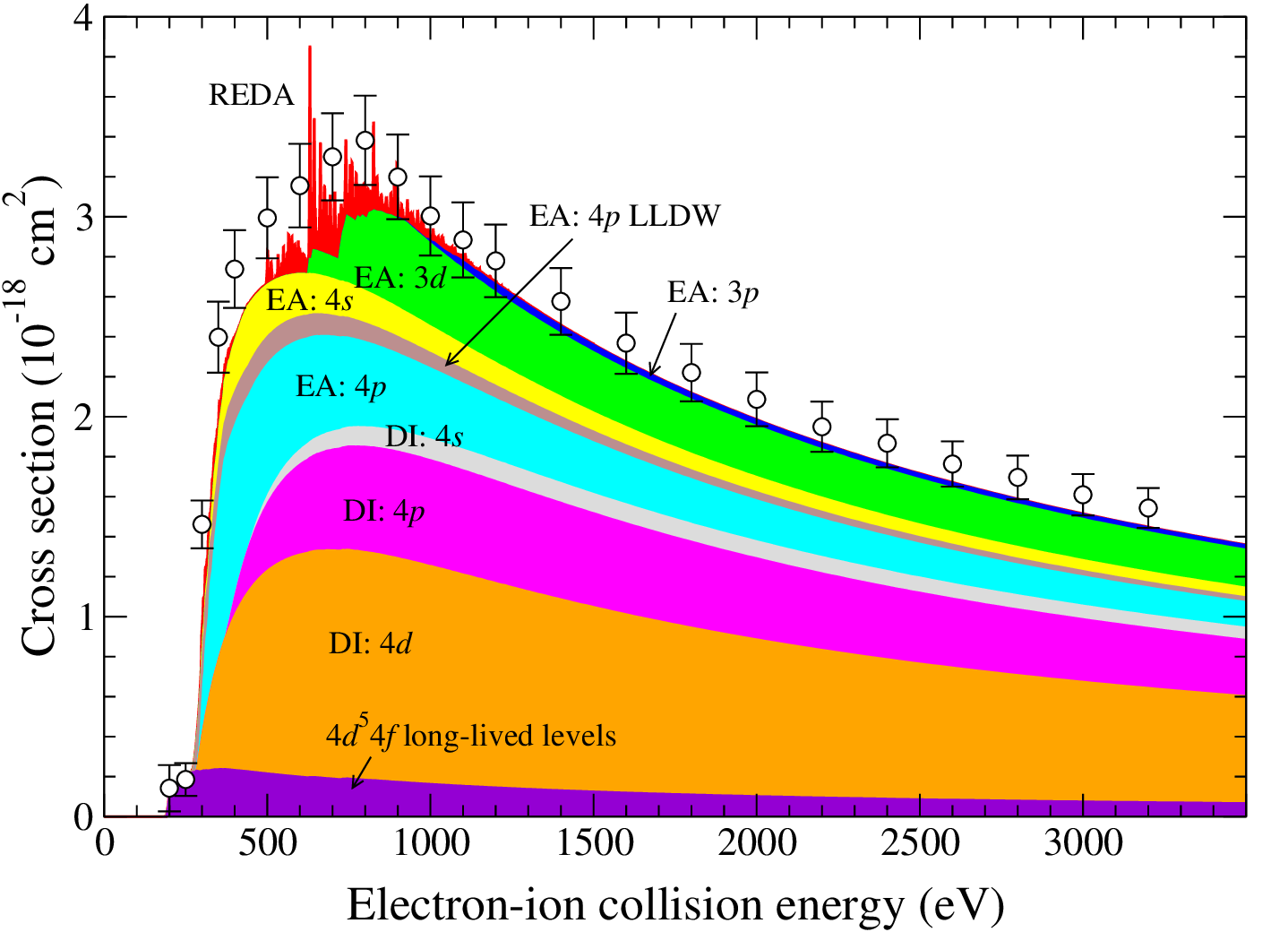}
	\caption{\label{xe12eisi-layers}
		EISI cross section of Xe$^{12+}$: Present experimental results (open circles with error bars) and hybrid SCADW/LLDW results assuming a mixture of the [Kr]$4d^{6}$ ground configuration (97\%) and the [Kr]$4d^{5}4f$ excited metastable configuration (3\%) for the primary ion-beam composition.}
\end{figure}

In addition to the improvements in our calculations discussed above, we also considered (weak) contributions from metastable excited primary ions, as was done already in our earlier work \cite{Borovik2015}. The presence of (long-lived) excited states in the primary ion beam employed in the experiment is responsible for the non-zero EISI cross section below the threshold for ground-level ionization of Xe$^{12+}$ at about 282 eV (Tab.~\ref{tab:thres}). These metastable states also contribute to all measured total cross sections in the entire experimental collision energy range. 
The metastable-ion fractions were adjusted in our computations to achieve the best possible match between the theoretical EISI cross sections and the corresponding experimental ones. This resulted in $4d^{6-x-1}4f$ metastable fractions of 3\%, 3\%, and 2\% for Xe$^{12+}$, Xe$^{13}$, and Xe$^{14+}$, respectively, largely in agreement with what was already obtained from our earlier CADW calculations \cite{Borovik2015}.

Figure \ref{xe12eisi-layers} compares the result of our improved calculation for EISI of Xe$^{12+}$ with the measured cross section. The theoretical cross section agrees with the measured data very well over the entire experimental energy range, notably now also in the range around and below the ground-level ionization threshold and up to the cross section maximum. The REDA contributions are decisive for the improved agreement at these energies. In Fig.~\ref{xe12eisi-layers}, the LLDW contribution to $4p$ EA is plotted separately. It makes up for almost 20\% of the entire $4p$ EA cross section.

\begin{figure}
	\includegraphics[width=\columnwidth]{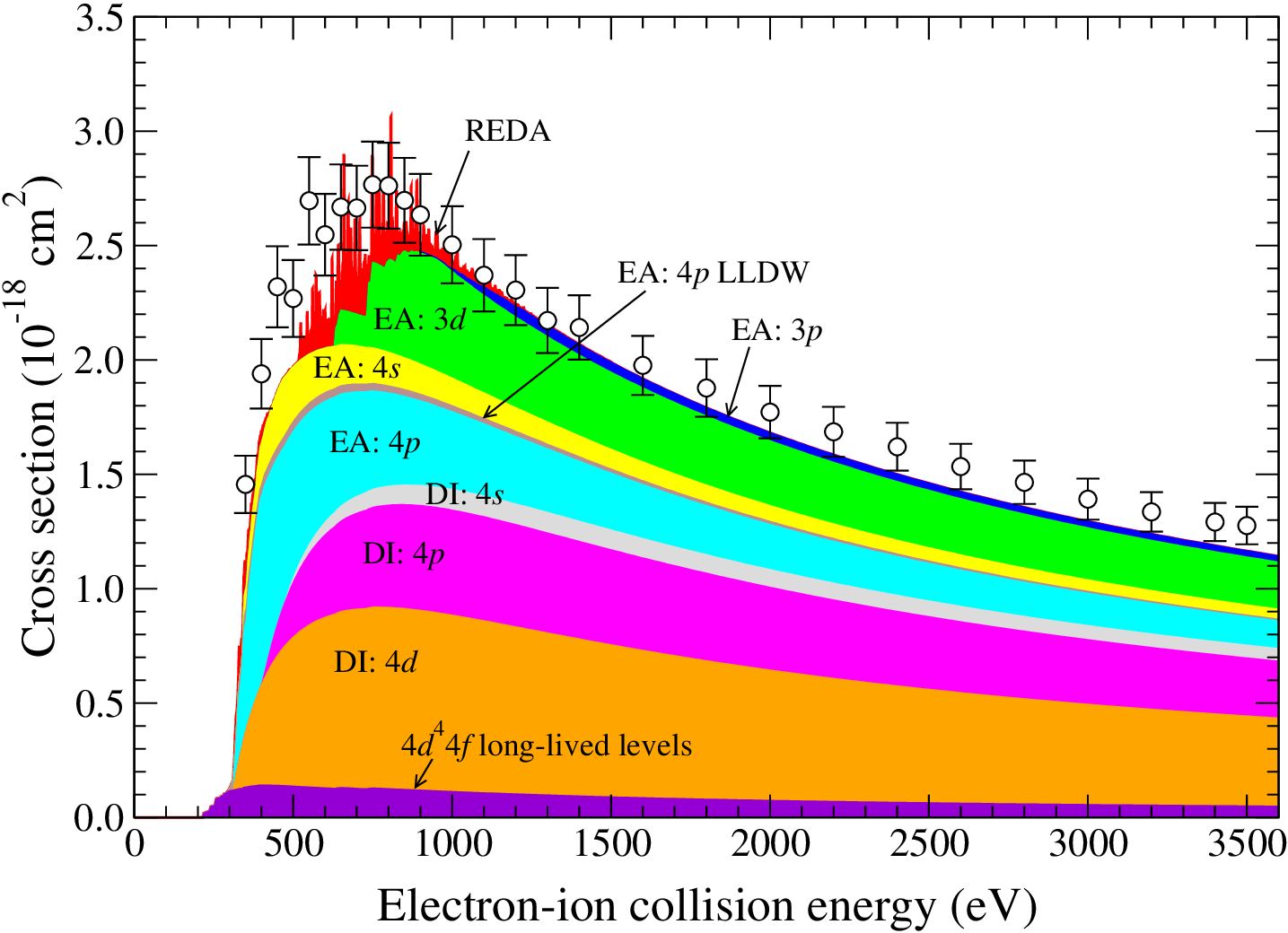}
	\caption{\label{xe13eisi-layers} 
		EISI cross section of Xe$^{13+}$: Present experimental results (open circles with error bars) and hybrid SCADW/LLDW results assuming a mixture of the [Kr]$4d^{5}$ ground configuration (97\%) and the [Kr]$4d^{4}4f$ excited metastable configuration (3\%)  for the primary ion-beam composition.}
\end{figure}

\begin{figure}
	\includegraphics[width=\columnwidth]{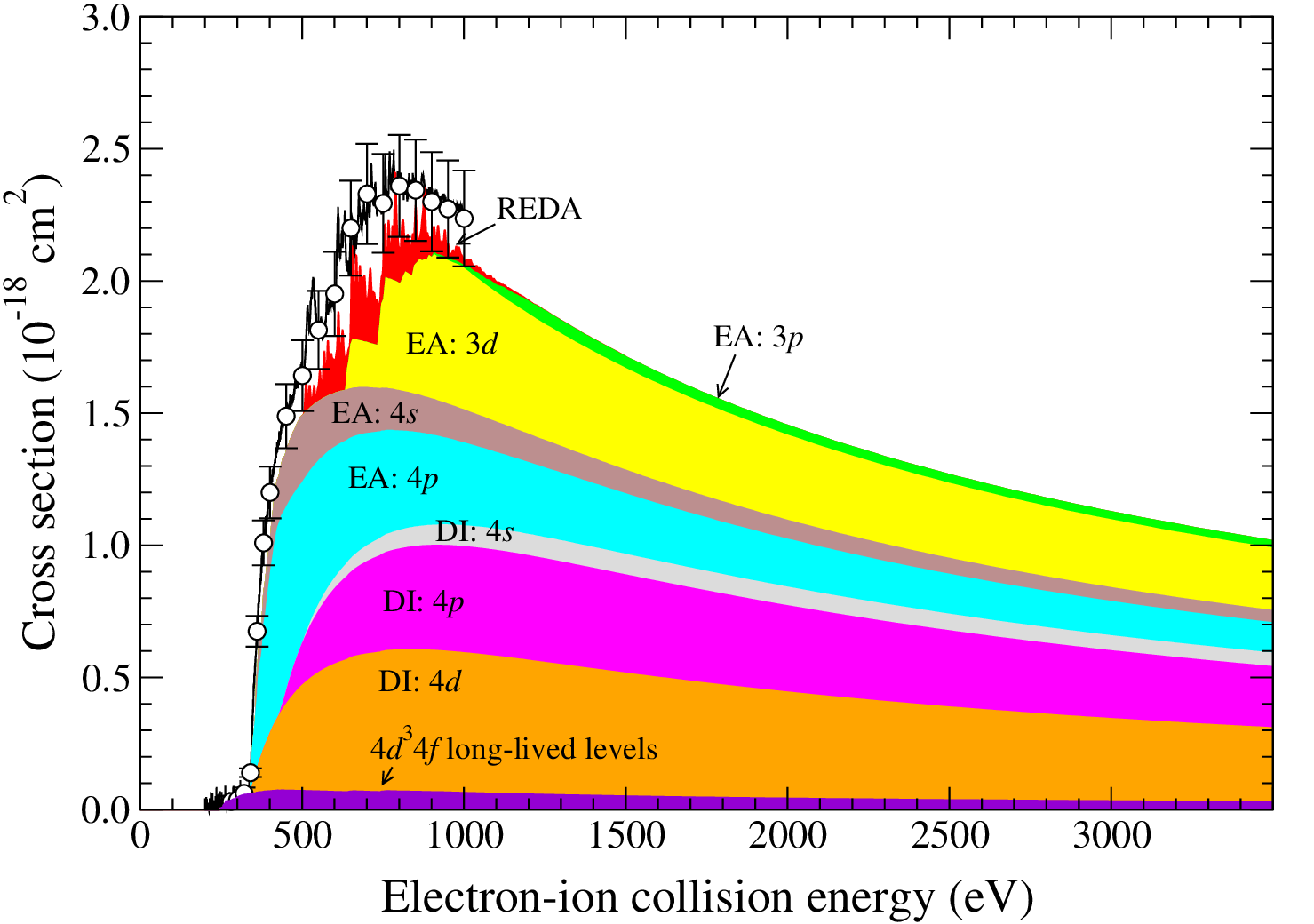}
	\caption{\label{xe14eisi-layers}{EISI cross section of Xe$^{14+}$: Experimental data from Ref.~\cite{Pindzola2013} (open circles with error bars) and present hybrid SCADW/LLDW results assuming a mixture of the [Kr]$4d^{4}$ ground configuration (98\%) and the [Kr]$4d^{3}4f$ excited metastable configuration (2\%) for the primary ion-beam composition. The thick black line represents experimental data \cite{Pindzola2013} that resulted from a fine energy scan and that were normalized to the absolute experimental data points \cite{Borovik2015}.}}
\end{figure}

Analogous to the approach described above, we also calculated the EISI cross sections of Xe$^{13+}$ and Xe$^{14+}$ ions. In Figs.~\ref{xe13eisi-layers} and \ref{xe14eisi-layers}, respectively, these are compared with the corresponding experimental data. In the case of Xe$^{14+}$ we compare with our previously published experimental data \cite{Pindzola2013,Borovik2015} since new EISI measurements have not been conducted for this ion. Also for Xe$^{13+}$ and Xe$^{14+}$ the inclusion of REDA contributions improves the agreement between theory and experiment decisively. The importance of the LLDW contribution to $4p$ EA obviously decreases with increasing charge state. For Xe$^{14+}$ it is almost negligible. This behavior does not follow a general rule as it is rather related to the specific atomic structures of the ions investigated. 

\subsection{Electron impact double ionization}

\begin{figure}
	\includegraphics[width=\columnwidth]{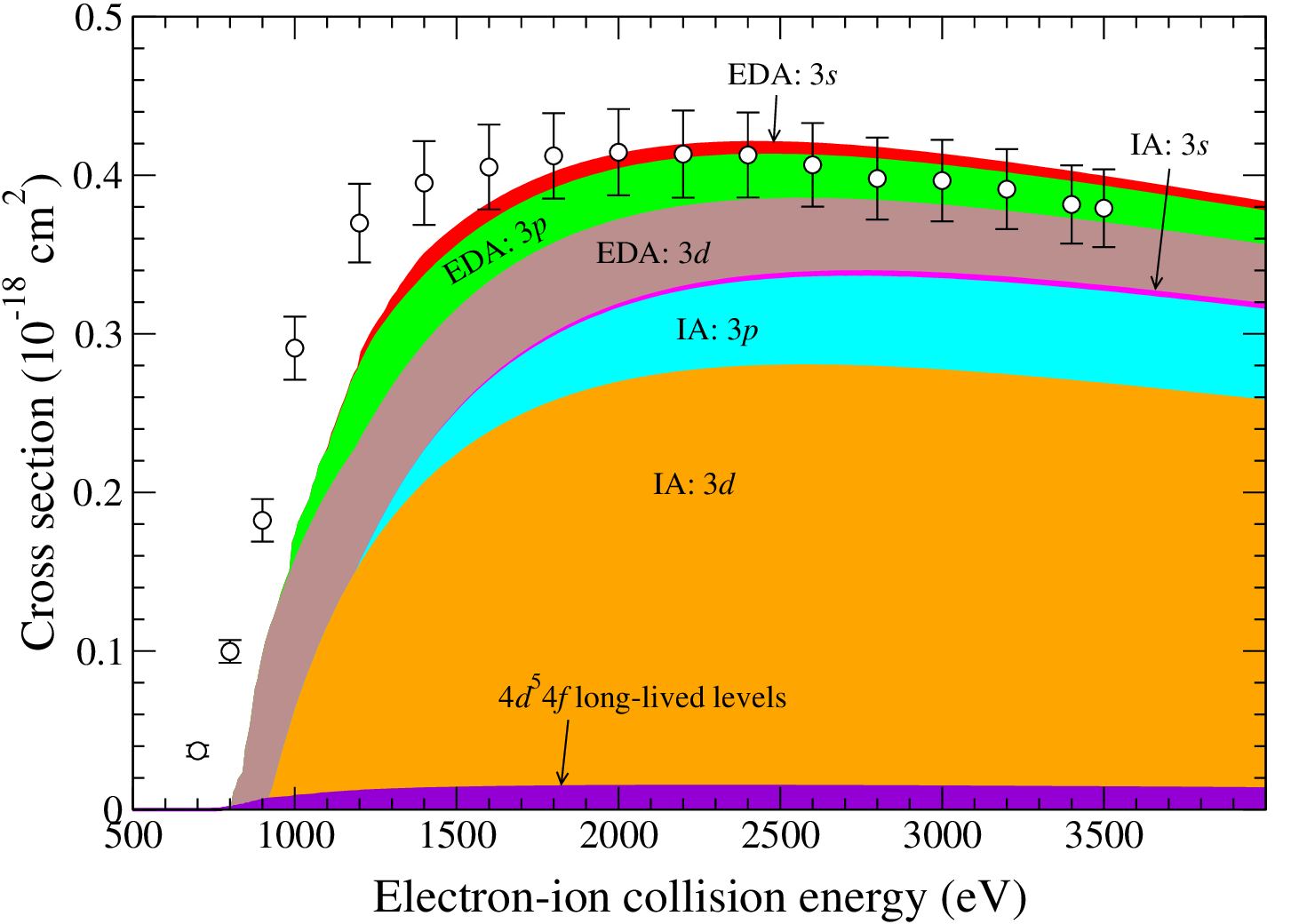}
	\caption{\label{xe12di} EIDI cross section of Xe$^{12+}$: Present experimental results (open circles with error bars) and SCADW results assuming a mixture of the [Kr]$4d^{6}$ ground configuration (97\%) and the [Kr]$4d^{5}4f$ excited metastable configuration (3\%).}
\end{figure}

\begin{figure}
	\includegraphics[width=\columnwidth]{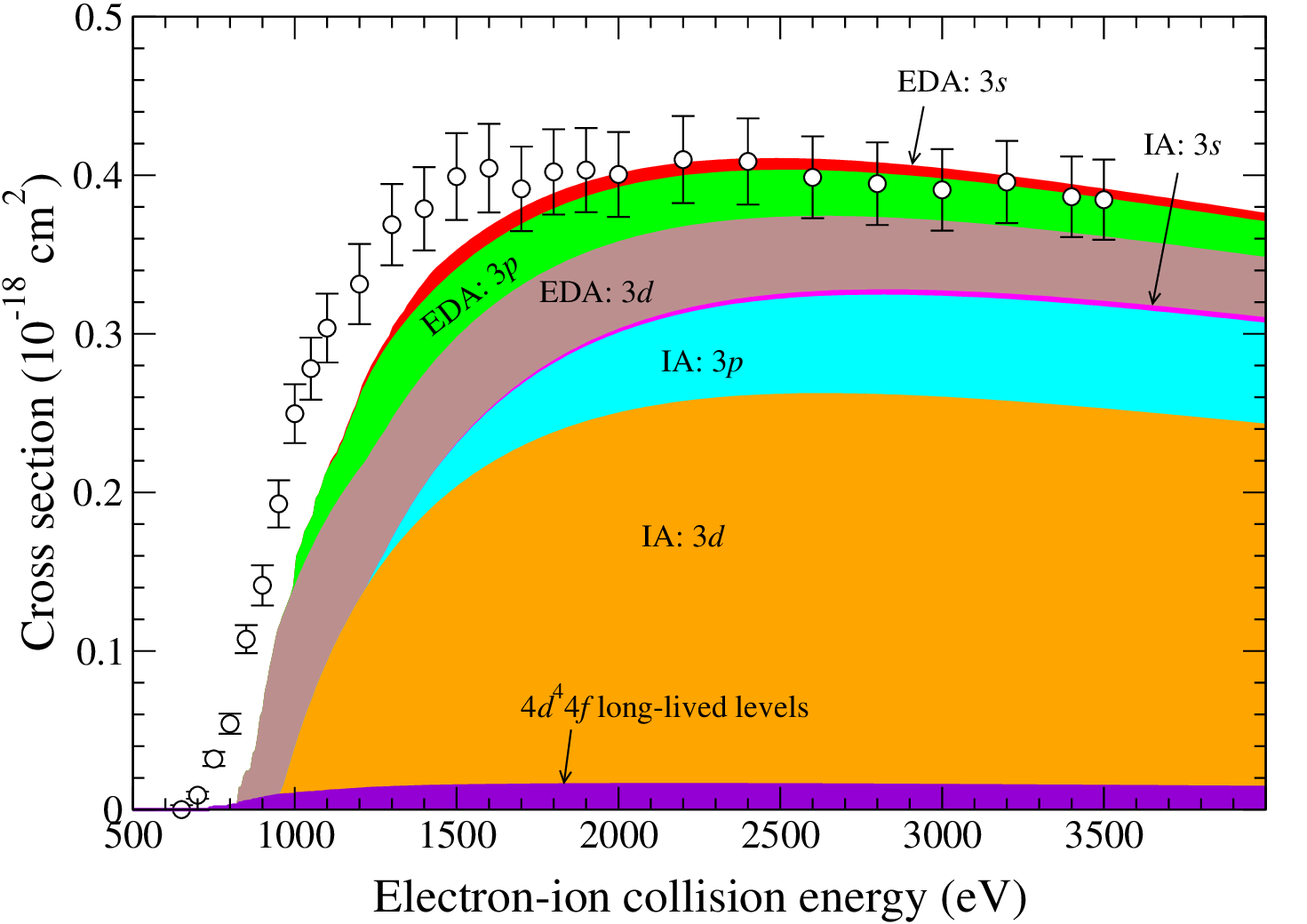}
	\caption{\label{xe13di} EIDI cross section of Xe$^{13+}$: Present experimental results (open circles with error bars) and SCADW results assuming a mixture of the [Kr]$4d^{5}$ ground configuration (97\%) and the [Kr]$4d^{4}4f$ excited metastable configuration (3\%).}
\end{figure}

\begin{figure}
	\includegraphics[width=\columnwidth]{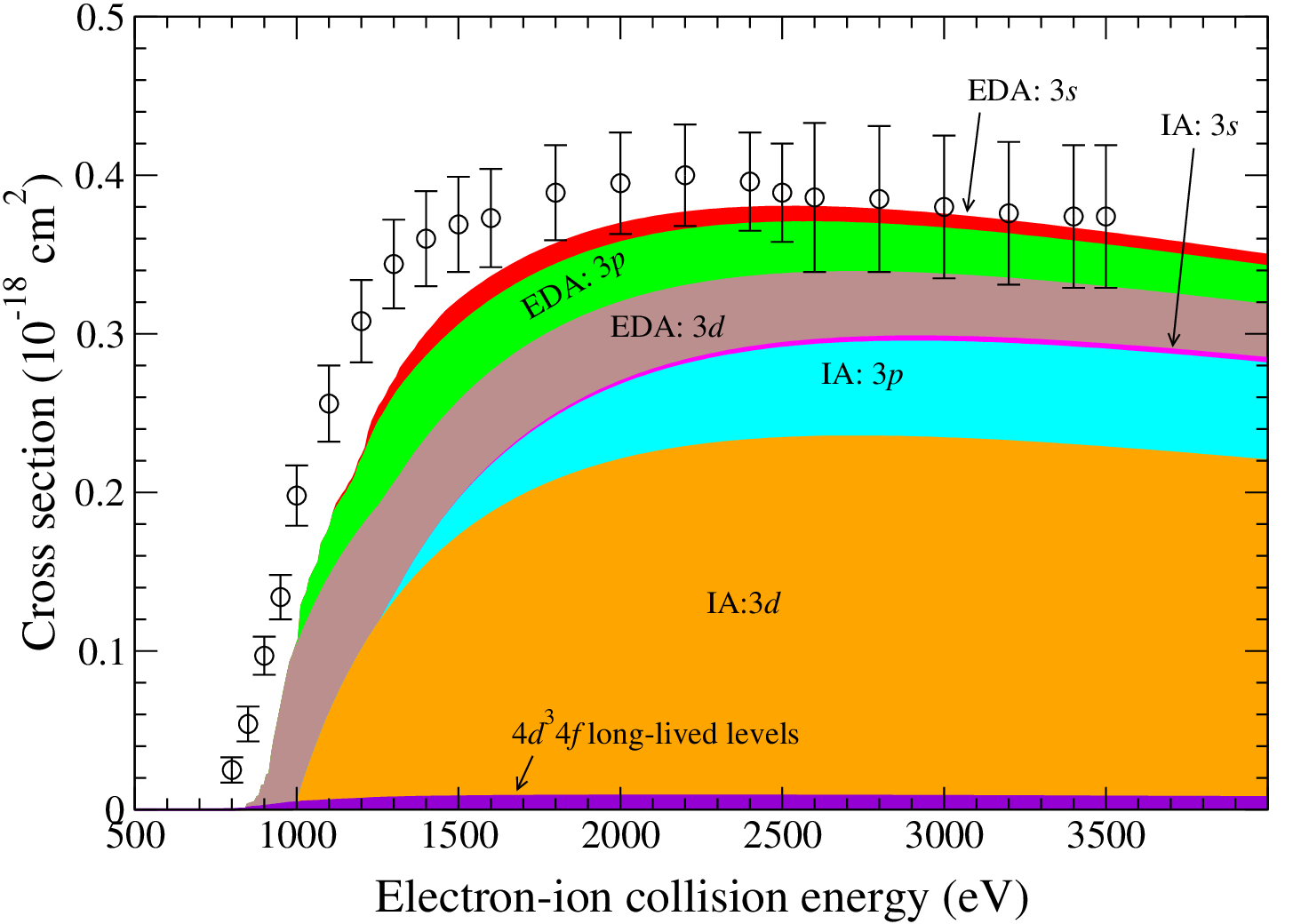}
	\caption{\label{xe14di} EIDI cross section of Xe$^{14+}$: Present experimental results (open circles with error bars) and SCADW results assuming a mixture of the [Kr]$4d^{4}$ ground configuration (98\%) and the [Kr]$4d^{3}4f$ excited metastable configuration (2\%).}
\end{figure}

Figures  \ref{xe12di}, \ref{xe13di}, and \ref{xe14di} present our measured and calculated EIDI cross sections for Xe$^{12+}$, Xe$^{13+}$, and Xe$^{14+}$, respectively. In general, the measured double-ionization cross sections rise less steeply than the respective single-ionization cross sections. None of the dominant IA or EDA  processes  straddles the threshold. Therefore, a detailed level-to-level treatment was not considered in our theoretical EIDI cross sections and all EDA and IA contributions (Eq.~\ref{eq:EIDI}) were calculated within the confines of the SCADW method. 

From our calculations we find that EDA and IA dominate the EIDI cross sections roughly below and above the cross section maxima, respectively.
The energetically lowest excitation channel into Xe$^{(12+x)+}$ configurations ($x=0-2$) above the double-ionization thresholds involves the excitation of a $3d$ electron to the $4d$ subshell. However, the ionized configurations Xe$^{(13+x)+}$($3d^{10}4s^{2}4p^{5}4d^{6-x}$) and Xe$^{(13+x)+}$($3d^{10}4s4p^{6}4d^{6-x}$), which can be reached from the $3d$$\to$$4d$ excited Xe$^{(12+x)+}$($3s^{2}3p^{6}3d^{9}4s^{2}4p^{6}4d^{7-x}$) configurations by a single Auger transition, are below the Xe$^{(13+x)+}$ single ionization thresholds. Therefore, $3d\to4d$ excitation contributes only to single but not to double ionization. EDA opens up only when a $3d$ electron is excited to the $5d$ or higher subshells, i.e., at energies much beyond the thresholds for double ionization. For example, the $3d$$\to$$5d$ exciation energy in Xe$^{12+}$ is $\sim$\,773~eV while the double ionization threshold occurs at about 593~eV (Tab.~\ref{tab:thres}). We conclude that EDA and IA cannot explain the experimental cross section in the vicinity of the double ionization threshold. At these energies, most probably the neglected (by us) DDI process (Eq.~\ref{eq:DDI}) contributes significantly to the total EIDI cross sections. In any case, our EDA and IA calculations yield total theoretical EIDI cross sections which agree with the experimental ones at high energies beyond about two times the double-ionization threshold. 

\section{\label{sec:conc}Conclusions}

Employing the electron-ion crossed-beams technique new experimental cross sections for electron-impact single-ionization (EISI) of Xe$^{12+}$ and Xe$^{13+}$ as well as for electron-impact double-ionization (EIDI) of Xe$^{12+}$, Xe$^{13+}$, and Xe$^{14+}$ were measured. A recently installed new electron gun \cite{Shi2003a,Ebinger2017,Doehring2020} allowed us to extend the  collision energy range from previously up to 1000~eV to now up to 3500~eV. The agreement of the new EISI data with our previously measured cross sections \cite{Pindzola2013,Borovik2015} testifies that the new gun operates as intended. With the new possibility to perform cross-section measurements at increased energies we can, in the future, extend our detailed EISI and EIDI studies to higher ion charge states and also to multiple ionization beyond double ionization. 

Different from earlier CADW calculations \cite{Borovik2011,Pindzola2013,Borovik2015} our present theoretical hybrid SCADW+LLDW approach leads to a satisfying agreement between theoretical and experimental EISI cross sections over the entire experimental energy range. The decisive ingredients for reaching this agreement were the detailed level-to-level calculations for excited configurations that straddle the ionization threshold and the inclusion of resonant-excitation double-autoionization (REDA) processes in the calculations. 

There is also good agreement between our experimental and theoretical EIDI cross sections for electron ion collision energies beyond two times the ionization threshold. At lower energies, the theoretical cross sections underestimate the experimental ones by large factors. We attribute this mainly to the neglect of direct double-ionization in our calculations. An even only approximate treatment of this ionization channel with three continuum electrons in the final state is beyond the scope of the present investigation. Nevertheless, we hope that our work stimulates future research in this direction.

\section*{Acknowledgments}

In this work the contributions from F.J. were supported by the State Key Laboratory of Laser Interaction with Matter (grant no. SKLLIM2008) and the National Natural Science Foundation of China (grant no. 
). Financial support by the German Federal Ministry of Education and Research (BMBF) within the \lq\lq{}Verbundforschung\rq\rq\ funding scheme (grant no.\ 05P19RGFA1) by the State of Hesse within the Cluster Project ELEMENTS is gratefully acknowledged.

\section*{Statement on data availability}

The data will be made available by the corresponding authors on reasonable requests.

\end{document}